\documentclass[conference]{IEEEtran}
\IEEEoverridecommandlockouts

\usepackage{amsmath,amssymb,amsfonts}
\usepackage{algorithmic}
\usepackage{booktabs}
\usepackage{graphicx}
\usepackage{siunitx}
\usepackage{adjustbox}

\usepackage{textcomp}
\usepackage[hyphens]{url}
\usepackage{hyperref}
\usepackage[nameinlink]{cleveref}
\usepackage{csquotes}
\usepackage{xcolor}
\usepackage{pifont}
\usepackage{tikz}
\usetikzlibrary{shapes,arrows,positioning,fit,backgrounds, calc}
\usepackage{xspace}
\usepackage[style=ieee, maxnames=3,citestyle=numeric-comp, minnames=3]{biblatex}
\hypersetup{
    colorlinks=true,        
    linkcolor=red,        
    citecolor=red,          
    urlcolor=blue           
}

\newcommand{\blackice}{\textsc{BlackIce}\xspace}

\newcommand{\filledcircle}{\tikz\fill[black] (0,0) circle (0.1cm);}
\newcommand{\halfcircle}{%
  \tikz[baseline={(0,-0.5ex)}]{
    \begin{scope}
      \clip (-0.1cm,-0.1cm) rectangle (0cm,0.1cm); 
      \fill[black] (0,0) circle (0.1cm);
    \end{scope}
    \draw (0,0) circle (0.1cm);
  }%
}
\newcommand{\emptycircle}{\tikz\draw (0,0) circle (0.1cm);}

\begin{document}

\title{\blackice: A Containerized Red Teaming Toolkit\\ for AI Security Testing}



\title{\blackice: A Containerized Red Teaming Toolkit\\ for AI Security Testing \thanks{*Accepted for presentation at the Conference on Applied Machine Learning in Information Security (CAMLIS) Red 2025.} \thanks{\textdagger~The authors thank Mrityunjay Gautam (Databricks) for proposing the initial idea and for his support of this project.}}

\author{
\IEEEauthorblockN{Caelin Kaplan, Alexander Warnecke, and Neil Archibald}
\IEEEauthorblockA{\textit{AI Red Team, Databricks}\\
\texttt{first.last@databricks.com}}
}

\maketitle

\begin{abstract}
AI models are being increasingly integrated into real-world systems, raising significant concerns about their safety and security. Consequently, AI red teaming has become essential for organizations to proactively identify and address vulnerabilities before they can be exploited by adversaries. While numerous AI red teaming tools currently exist, practitioners face challenges in selecting the most appropriate tools from a rapidly expanding landscape, as well as managing complex and frequently conflicting software dependencies across isolated projects. Given these challenges and the relatively small number of organizations with dedicated AI red teams, there is a strong need to lower barriers to entry and establish a standardized environment that simplifies the setup and execution of comprehensive AI model assessments.

Inspired by Kali Linux’s role in traditional penetration testing, we introduce \blackice, an open-source containerized toolkit designed for red teaming Large Language Models (LLMs) and classical machine learning (ML) models. \blackice provides a reproducible, version-pinned Docker image that bundles 14 carefully selected open-source tools for Responsible AI and Security testing, all accessible via a unified command-line interface. With this setup, initiating red team assessments is as straightforward as launching a container, either locally or using a cloud platform. Additionally, the image's modular architecture facilitates community-driven extensions, allowing users to easily adapt or expand the toolkit as new threats emerge. In this paper, we describe the architecture of the container image, the process used for selecting tools, and the types of evaluations they support.
\end{abstract}


\section{Introduction}
As AI systems become increasingly integrated into critical workflows and consumer products, AI red teaming has emerged as an essential practice to identify and mitigate vulnerabilities. These vulnerabilities can manifest at the model level, such as jailbreak attacks~\cite{ZouWanKol+23, HuaGupXia+24, WeiHagSte23} that circumvent safety mechanisms, or at the system level, where adversaries exploit deployment contexts, for example, indirect prompt injections embedded within emails processed by AI assistants~\cite{GreAbdMish+23,LiuDenLi+23}. Despite the growing recognition of AI red teaming’s importance, effectively conducting such assessments remains challenging, as existing tools each have their own unique setup procedures and typically require separate runtime environments due to conflicting dependencies. While possible, managing many independent environments is often time-consuming, error-prone, and difficult to scale, particularly across diverse platforms or cloud-based infrastructure. Moreover, the absence of a single, standardized environment complicates the reproducibility of evaluations across teams and organizations.

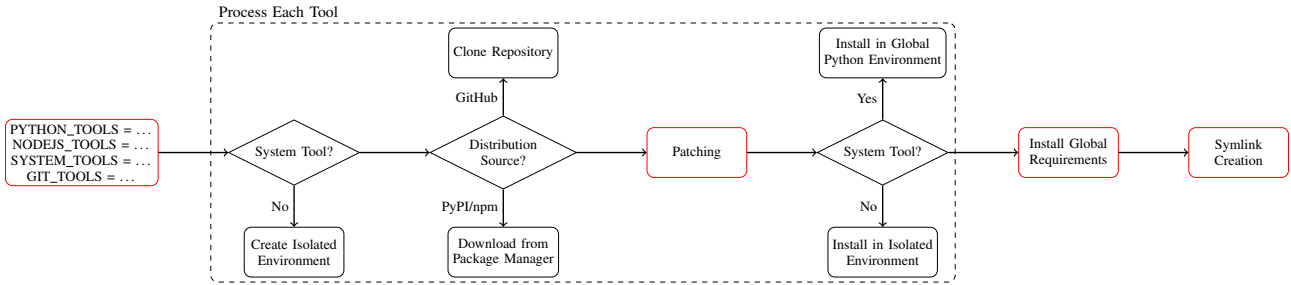
\begin{figure*}
    \centering
    \begin{adjustbox}{width=0.95  \textwidth}
        \def\YOffset{2.0cm}
\begin{tikzpicture}[
    node distance=1.4cm,
    box/.style={rectangle, draw, rounded corners, minimum width=2cm, minimum height=1cm, align=center},
    decision/.style={diamond, draw, align=center, aspect=2},
    arrow/.style={->, thick},
    note/.style={rectangle, draw, dashed, fill=gray!10, text width=3cm},
    tool/.style={rectangle, draw, fill=blue!10, text width=3cm},
    process/.style={rectangle, draw, fill=green!10, text width=2.5cm},
    process2/.style={rectangle, draw, fill=green!10, text width=3.8cm}
]
\footnotesize 
\node[box, draw=red] (get-tool) {PYTHON\_TOOLS = \dots \\ NODEJS\_TOOLS = \dots \\ SYSTEM\_TOOLS = \dots \\ GIT\_TOOLS = \dots};
\node[decision, right=of get-tool] (venv) {System Tool?};
\coordinate (venv-center) at (venv);
\node[box] at ([yshift=-\YOffset]venv-center) (create-venv1) {Create Isolated \\ Environment};

\node[decision, right=of venv] (tool-type) {Distribution\\Source?};
\coordinate (tooltype-center) at (tool-type);
\node[box] at ([yshift=\YOffset]tooltype-center) (git-tool) {Clone Repository};
\node[box] at ([yshift=-\YOffset]tooltype-center) (pypi-tool) {Download from \\ Package Manager};

\node[box, right=of tool-type, draw=red] (patching) {Patching};

\node[decision, right=of patching] (is-system) {System Tool?};
\coordinate (issystem-center) at (is-system);
\node[box] at ([yshift=\YOffset]issystem-center) (system-install) {%
  Install in Global\\
  Python Environment
};
\node[box] at ([yshift=-\YOffset]issystem-center) (final-systool) {%
  Install in Isolated\\
  Environment
};

\node[box, draw=red, right=of is-system] (final-check) {Install Global \\ Requirements};
\coordinate (decision-center) at (final-check);

\node[box, right=of final-check, draw=red] (symlinks) {Symlink \\ Creation};

\draw[arrow] (get-tool) -- (venv);
\draw[arrow] (venv) -- (tool-type);
\draw[arrow] (venv) -- node[left] {No} (create-venv1);
\draw[arrow] (tool-type) -- node[left] {GitHub} (git-tool);
\draw[arrow] (tool-type) -- node[left] {PyPI/npm} (pypi-tool);
\draw[arrow] (tool-type) -- (patching);
\draw[arrow] (patching) -- (is-system);

\draw[arrow] (is-system) -- node[left] {Yes} (system-install);
\draw[arrow] (is-system) -- node[left] {No} (final-systool);
\draw[arrow] (is-system) -- (final-check);
\draw[arrow] (final-check) -- (symlinks);


\begin{scope}[on background layer]
\node[fit=(venv) (git-tool) (final-systool), 
      draw=black,
      fill=white,
      rounded corners,
      dashed,
      inner xsep=0.35cm,
      name=loopbox] {};

  \node[anchor=north west, font=\small, xshift=2pt, yshift=12pt] 
        at (loopbox.north west) {Process Each Tool}; 
\end{scope}

\end{tikzpicture}
    \end{adjustbox}
    \caption{Flowchart illustrating the Docker build process for the \blackice container image. Boxes with red borders highlight the steps where users directly interact with the build process to integrate new tools.}
    \label{fig:flowchart}
\end{figure*}

In traditional penetration testing, similar challenges have been effectively addressed through the widespread adoption of Kali Linux, a Linux distribution preconfigured with a comprehensive suite of security tools. Kali Linux has become a standard in the security community by simplifying environment setup and bundling essential utilities, enabling practitioners to focus on vulnerability assessments rather than managing complex software configurations and dependencies. Inspired by Kali Linux's success, we introduce \blackice, an open-source toolkit that consolidates leading AI red teaming tools into a unified, reproducible, and portable container image\footnote{\url{https://hub.docker.com/r/databricksruntime/blackice}}; the Docker build repository is available at \mbox{\url{https://github.com/databricks/containers}}. The modular architecture and carefully curated tool selection of \blackice enable both novice and expert practitioners to effectively perform AI red team assessments, while also facilitating straightforward, community-driven extensions. An overview of the included tools is provided in~\Cref{tab:llm-tools}, and the corresponding Docker build process is illustrated in~\Cref{fig:flowchart}.

\begin{table}[ht]
\centering
\begin{tabular}{llS[table-format=4.0]cl}
\toprule
\textbf{Tool} & \textbf{Organization} & \textbf{Stars} & \textbf{Type} & \textbf{Source} \\
\midrule
Eval Harness~\cite{eval-harness}  & Eleuther AI & 10300 & \filledcircle & GitHub \\ 
Promptfoo~\cite{promptfoo}              & Promptfoo & 8600 & \filledcircle & npm \\
CleverHans~\cite{cleverhans}              & CleverHans Lab & 6400 & \emptycircle & GitHub \\
Garak~\cite{garak}                      & NVIDIA & 6100 & \filledcircle & PyPI \\
ART~\cite{art}              & IBM & 5600 & \emptycircle & PyPI \\
Giskard~\cite{giskard}                  & Giskard & 4900 & \halfcircle & PyPI \\
CyberSecEval~\cite{purplellama}          & Meta & 3800 & \filledcircle & GitHub \\
PyRIT~\cite{pyrit}                      & Microsoft & 2900 & \emptycircle & PyPI \\
EasyEdit~\cite{easyedit}                & ZJUNLP & 2600 & \emptycircle & GitHub \\
Promptmap~\cite{promptmap}              & - & 1000 & \filledcircle & GitHub \\
FuzzyAI~\cite{fuzzyai}                  & CyberArk & 800 & \filledcircle & GitHub \\
Fickling~\cite{fickling}                & Trail of Bits & 560 & \halfcircle & PyPI \\
Rigging~\cite{rigging}                  & Dreadnode & 380 & \emptycircle & PyPI \\
Judges~\cite{judges}                    & Quotient AI & 290 & \halfcircle & PyPI \\
\bottomrule
\end{tabular}
\vspace{1mm}
\caption{Overview of \blackice{} tools, including organization (if applicable), GitHub stars (nearest hundred), tool type (Static~\protect\filledcircle{} or Dynamic~\protect\emptycircle{}), and distribution source.}
\label{tab:llm-tools}
\end{table}

\section{Docker Image Architecture}
\label{sect:sec2}
AI red teaming tools vary significantly in their usage, dependencies, and distribution methods, posing unique challenges when combining them effectively into a unified workflow. While managing separate runtime environments for each tool during an assessment is possible, this approach has three major drawbacks: (1) it significantly increases operational complexity due to repetitive configuration and frequent environment switching, (2) it restricts interoperability between tools by isolating their environments, and (3) it creates challenges for cloud-based managed notebook services, which typically provide only a single Python interpreter per kernel.

\paragraph{Static vs. Dynamic Tools}
To enable seamless use of all tools within the same container environment, we organize them into two distinct categories based on usage:
\begin{itemize}
\item \textbf{Static Tools:} Evaluate AI models using straightforward command-line interfaces, requiring minimal programming knowledge. While easy to use, static tools offer limited flexibility for customization and integration.
\item \textbf{Dynamic Tools:} Offer similar evaluation capabilities but additionally support advanced Python-based customization, enabling users to write code for custom attacks.
\end{itemize}
Within the container image, static tools are installed in individual, isolated Python virtual environments (or separate Node.js projects), each with its own dependencies, and can be executed immediately via symlinks to custom CLI scripts. Alternatively, dynamic tools are installed directly into the global Python environment, with dependency conflicts managed through a \texttt{global\_requirements.txt} file.

\paragraph{Automated Installation}
The Docker build process automates the installation and configuration of each tool according to its category and distribution source. At the top of the Dockerfile, users specify tool names along with their pinned versions (defined either by package version or Git commit hash) in predefined configuration lists. By default, tools listed under \texttt{PYTHON\_TOOLS} are installed from PyPI into isolated Python virtual environments. Users can override this default behavior by adding tools to \texttt{SYSTEM\_TOOLS}, which installs them into the global Python environment, or to \texttt{GIT\_TOOLS}, which clones and installs them directly from their Git repositories. \texttt{NODEJS\_TOOLS} are always installed via npm into separate project directories.

\paragraph{Customization via Patching}
In certain scenarios, users may need to extend or modify the functionality of existing tools. For example, an assessment might require implementing a custom client to query a model endpoint not supported by default, or enhancing a tool like CyberSecEval~\cite{purplellama} to enable providing the LLM with a security-oriented system prompt during benchmark evaluations. To simplify these modifications, the \blackice build process includes a structured patching mechanism. Users can add source-level patches as simple \texttt{.diff} files or custom Python modules to a designated directory (\texttt{/patches/<tool>}), which are automatically applied before the tool is installed.

\paragraph{Community Extensibility}
The modular build design enables users to easily add new tools by appending them to the configuration lists defined at the top of the Dockerfile (e.g., \texttt{PYTHON\_TOOLS}, \texttt{SYSTEM\_TOOLS}, \texttt{GIT\_TOOLS}). After doing so, static tools are integrated by creating a minimal wrapper script in the \texttt{/cli\_scripts/<tool>} directory that includes a shebang pointing to the tool’s isolated Python interpreter and invokes its run command with the provided arguments. Custom static tools can be added in the same way by placing a self-contained CLI script in \texttt{/cli\_scripts/<tool>}. For example, we include \texttt{biasforge}, a custom CLI tool that evaluates bias in language models by generating synthetic prompts based on a specified evaluation objective, querying a target model, and assessing the outputs using a structured judgment schema. Dynamic tools, in contrast, may require updating the \texttt{global\_requirements.txt} file to resolve any newly introduced dependency conflicts. Once the image is rebuilt, all new tools become immediately available via the CLI, with no additional setup required.

\section{Tool Selection and Coverage}
\label{sect:sec3}
\begin{table*}[ht]
\centering
\small
\renewcommand{\arraystretch}{1.4} 
\begin{tabular}{p{0.30\textwidth} p{0.34\textwidth} p{0.29\textwidth}}
\toprule
\textbf{BlackIce Capability} & \textbf{MITRE ATLAS} & \textbf{Databricks AI Security Framework (DASF)} \\
\midrule

Prompt-injection and jailbreak testing of LLMs &
AML.T0051 LLM Prompt Injection; \newline
AML.T0054 LLM Jailbreak; \newline
AML.T0056 LLM Meta Prompt Extraction &
9.1 Prompt inject; \newline
9.12 LLM jailbreak \\

\midrule
Indirect prompt injection via untrusted content (e.g., RAG/email) &
AML.T0051 LLM Prompt Injection [Indirect] &
9.9 Input resource control \\

\midrule
LLM data leakage testing &
AML.T0057 LLM Data Leakage &
10.6 Sensitive data output from a model \\

\midrule
Hallucination stress-testing and detection &
AML.T0062 Discover LLM Hallucinations &
9.8 LLM hallucinations \\

\midrule
Adversarial example generation and evasion testing (CV/ML) &
AML.T0015 Evade ML Model; \newline
AML.T0043 Craft Adversarial Data &
10.5 Black box attacks \\

\midrule
Supply-chain and artifact safety scanning (e.g., malicious pickles) &
AML.T0010 AI Supply Chain Compromise; \newline
AML.T0011 Unsafe AI Artifacts &
7.3 ML supply chain vulnerabilities \\

\bottomrule
\end{tabular}
\vspace{1mm}
\caption{Mapping of \blackice{} capabilities to MITRE ATLAS and the Databricks AI Security Framework (DASF).}
\label{tab:capability-mapping}
\end{table*}

To ensure comprehensive coverage, we first identified critical threat domains based on technical reports and red teaming guidelines published by industry-leading organizations~\cite{GoogleReport,BulMinCha+25,attackAtlas,ciscoReport,BAHReport,AnthropicReport,anthropicBlog}. These reports provided valuable insights into common techniques and best practices used in AI red team assessments. Second, we examined \emph{system cards} (e.g.,~\cite{SystemCardGoogle, SystemCardClaude, SystemCardOpenAI}), structured documents transparently detailing organizations' internal red teaming processes and findings. Finally, we reviewed outcomes from \emph{red teaming competitions} (e.g.,~\cite{msPlaygroundLabs, defconChallenge, dreadnodeCrucible}), where participants attack LLMs in controlled environments designed to reflect real-world adversarial scenarios, similar to “capture the flag” challenges.
Guided by this analysis and discussions with industry AI security practitioners, we selected widely adopted open-source tools, including those developed by established AI security teams (e.g.,~\cite{garak,pyrit,purplellama}), AI security startups (e.g., ~\cite{promptfoo,giskard}), academic researchers~\cite{cleverhans,easyedit}, and independent developers~\cite{promptmap}. The selected tools were then evaluated for their collective coverage of major AI security risk categories by mapping the capabilities of \blackice to MITRE ATLAS~\cite{mitreatlas} and the Databricks AI Security Framework (DASF)~\cite{dasf}. Table~\ref{tab:capability-mapping} highlights that \blackice provides comprehensive coverage across domains such as prompt injection, data leakage, hallucination detection, and supply-chain integrity.

\paragraph{Static Tool Capabilities}
Although there is a certain degree of functional overlap among static tools, each provides distinct testing capabilities and excels in particular areas. Evaluation Harness~\cite{eval-harness} offers over 60 academic benchmarks targeting skills such as reasoning, reading comprehension, and mathematics; Promptfoo~\cite{promptfoo} generates context-specific attacks and can evaluate vulnerabilities according to frameworks such as the OWASP Top 10 for LLMs~\cite{owasptop10}; Garak~\cite{garak} leverages predefined prompt sets to assess model-level vulnerabilities including bias, toxicity, jailbreaks, and hallucinations; Giskard~\cite{giskard} provides an extensible automated testing platform supporting integration with multiple LLM providers;\footnote{Although Giskard did not originally provide a CLI, we implemented one to classify it as a static tool, given its strong static evaluation capabilities.} CyberSecEval~\cite{purplellama} targets security-specific issues such as unsafe code generation and malware synthesis through predefined corpuses; and Fickling~\cite{fickling} scans Python pickle files for malicious payloads. 

\paragraph{Dynamic Tool Capabilities}
For advanced users, dynamic tools enable conducting more customized assessments: PyRIT~\cite{pyrit} is a Python framework allowing users to configure red teaming workflows using components such as Prompt Targets, Executors, Scorers, and Converters, facilitating comprehensive and easily extendable testing scenarios; EasyEdit~\cite{easyedit} enables direct manipulation of a model’s internal representations, allowing users to alter stored knowledge; and Rigging~\cite{rigging} is a framework for orchestrating type-safe LLM workflows, ideal for automating structured tasks in AI security assessments.\footnote{Technically, EasyEdit, Cleverhans, and ART each qualify as dynamic tools under our definition, meaning they should be installed in the system-level Python environment. However, due to significant dependency conflicts between their legacy requirements and modern LLM evaluation tools (e.g., PyRIT), we instead treat them as static tools and install each in its own Python virtual environment.}

\section{Conclusion}
\blackice is released as an open-source Docker image along with its build repository, providing a standardized execution environment that greatly reduces the effort required to conduct comprehensive AI red teaming assessments. The image includes a curated selection of static and dynamic tools, collectively addressing a broad range of vulnerabilities in both LLMs and classical ML models. Through its modular design and straightforward extensibility, we aim for \blackice to foster community-driven enhancements and encourage responsible AI deployment practices.

\begingroup
\renewcommand*{\bibfont}{\footnotesize}  
\setlength{\bibitemsep}{0pt}             

\printbibliography
\endgroup

\vspace{12pt}

\end{document}